\documentclass[
twocolumn,
superscriptaddress,
 amsmath,amssymb,
 aps,prl
]{revtex4}

\usepackage{graphicx}% Include figure files
\usepackage{dcolumn}% Align table columns on decimal point
\usepackage{bm}
\usepackage{wrapfig}
\usepackage{float}% bold math
\begin{document}

\preprint{APS/123-QED}

\title{Geometric Mechanics of Curved Crease Origami}

\author{Marcelo A. Dias}
\email{madias@physics.umass.edu}
\affiliation{Department of Physics, University of Massachusetts Amherst, Amherst, Massachusetts 01002}
\author{Levi H. Dudte}
\email{ldudte@seas.harvard.edu}
\affiliation{Engineering and Applied Sciences, Harvard University, Cambridge, Massachusetts 02138}
\author{L. Mahadevan}
\email{lm@seas.harvard.edu}
\affiliation{Physics and Engineering and Applied Sciences, Harvard University, Cambridge, Massachusetts 02138}
\author{Christian D. Santangelo}
\email{csantang@physics.umass.edu}
\affiliation{Department of Physics, University of Massachusetts Amherst, Amherst, Massachusetts 01002}

\begin{abstract}
Folding a sheet of paper along a curve can lead to structures seen in decorative art and utilitarian packing boxes.  Here we present a theory for the simplest such structure: an annular circular strip that is folded along a central circular curve to form a three-dimensional buckled structure driven by geometrical frustration. We quantify this shape in terms of the radius of the circle, the dihedral angle of the fold and the mechanical properties of the sheet of paper and the fold itself. When the sheet is isometrically deformed everywhere except along the fold itself, stiff folds result in creases with constant curvature and oscillatory torsion.  However, relatively softer folds inherit the broken symmetry of the buckled shape with oscillatory curvature and torsion. Our asymptotic analysis of the isometrically deformed state is corroborated by numerical simulations which allow us to generalize our analysis to study multiply folded structures.
\end{abstract}

\date{\today}

\maketitle

Over the last few decades, origami has evolved from an art form into a scientific discipline \cite{DemaineOrourke2009,demaine1}, made possible in large part by methods for the mathematical analysis of folded surfaces. Though there is artistic and technological precedent for folding paper along curves
\cite{Huffman76,DuncanDuncan82,Fuchs and Tabachnikov99,PottmannWallner2001,kilianfold}, as seen in Japanese paper boxes, and the ubiquitous McDonald's fries box, how these geometrical structures get their physical shape is poorly understood. In this letter, we consider the geometric mechanics of the simplest case of a curved fold: an annular elastic sheet folded along a closed, circular curve which buckles out of the plane to resolve a fundamental incompatibility between the folded geometry and the ensuing mechanical stresses.

\begin{figure}[b]
\includegraphics[width=3.3in]{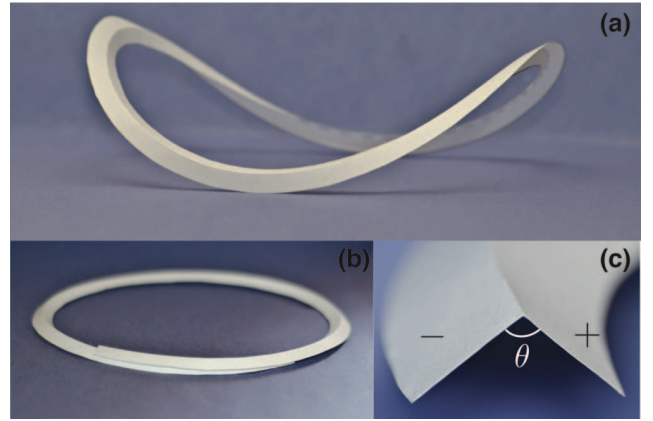}
\caption{\label{fig:fold2} (color online) A photograph of the model built by cutting a flat annulus of width $2w$ from a flat sheet of paper with central circle of radius $r$. (a) Folding along its center line buckles the structure out-of-plane. However, if we cut the annulus, (b), the structure collapses to an overlapping planar state with curvature given by equation (\ref{eq:constr1}). (c) The inset shows a cross section of the fold, where the right and left planes, $\mathbf{X}_+$ and $\mathbf{X}_-$, define the dihedral angle $\theta$.}
\end{figure}

Qualitative experiments with a complete circular annulus of paper having a concentric, circular crease show that folding buckles the crease into a saddle (Fig. \ref{fig:fold2}a), while the same crease along a cut annulus remains planar (Fig. \ref{fig:fold2}b). This behavior is a consequence of a fundamental incompatibility between the geometry of the fold and the stretching elasticity of the sheet. As we will see, and it is apparent in Fig. \ref{fig:fold2}b, the sheet responds to folding by wrapping around itself to eliminate in-plane mechanical stresses. The closed annulus, on the other hand, can expel these stresses by buckling.  In the limit of where the thickness of the sheet is much smaller than the width, which is itself smaller than the length of the crease, the shape that arises is a balance between the bending energy of the sheets on either side of the crease, the energy at the crease itself, and the geometrical constraints arising from the sheet's closed topology.
%Added
Topological constraints are also crucial to shape and mechanics in the small deformations of shells \cite{audoly} and non-Euclidean plates \cite{santangelo}.
%%%%%%%%%%%%%%%%%%%%%%%%%
%%%%%%%%%%%%%%%%%%%%%%%%%%%%%%
%%%

We consider an annulus of uniform thickness $t$, width $2w$ folded along a central circular crease of radius $r$ ($t \ll w < 2 \pi r$). In the deformed state, the crease is a space curve parametrized by arc length $s$, with curvature $\kappa(s)$ and torsion $\tau(s)$, and the surfaces on either side of it come together at a finite fold (or dihedral) angle $\theta(s)$. Assuming isometric deformations away from the crease, the mid-surface of the sheet on either side of the crease is developable. Then any point on it can be characterized in terms of a set of coordinates $(s, v)$, corresponding to the arc length and the generators of the developable on the inside and outside of the crease, $\mathbf{g}_\pm$ (Fig. \ref{fig:fold2}c), with  the coordinates: $\mathbf{X}_\pm(s,v) = \mathbf{X}(s,0) + v \mathbf{g}_\pm(s)$. For developables, the generators must also satisfy the condition that $\mathbf{g}_\pm(s) \times d\mathbf{g}_{\pm}(s)/ds$ is perpendicular to the crease \cite{doCarmo76}. Since folding does not induce in-plane strains, the projection of the crease curvature onto the tangent plane on either side of the sheet must remain $1/r$. This leads to two geometrical conditions \cite{Fuchs and Tabachnikov99} that relate the dihedral angle of the crease to its spatial curvature and the angle of the generators of the developable surface on either side of it . They read 
\begin{eqnarray}\label{eq:constr1}
 \sin \left(\frac{\theta}{2}\right) &=& \frac{1}{\kappa},\\
\cot \gamma_{\pm}&=&- \frac{1}{2} \left(2 \tau\pm r \frac{d \theta}{ds} \right) \tan\left(\frac{\theta}{2}\right),
\end{eqnarray}
where $\kappa/r$ and $\tau/r$ are the curvature and torsion of the crease respectively, $\gamma_\pm$ is the angle between the unit tangent vector of the crease $d \mathbf{X}(s)/ds$ and the generator. We see that $\kappa(s) \ge 1$, with equality only when $\theta=\pi$.  For a circular crease concentric with a circular annulus of constant dimensionless half-width $\omega=w/r$, we find
\begin{equation}\label{eq:vmax}
v^{max}_\pm(\xi) = \pm \sin \gamma_\pm(\xi) \mp \sqrt{\omega^2 \mp 2 \omega + \sin^2 \gamma_\pm(\xi)}
\end{equation}
to be the dimensionless distance to the boundary along a generator leaving the crease from a point labeled by the dimensionless arc length $\xi=s/r$.

The energy of the sheet is the sum of the energy of deforming the sheet on either side of the crease and that of the fold that connects them. Since the creased folded surface is piecewise developable, the energy per unit surface is proportional to the square of the mean curvature \cite{Love}. The mean curvature on either side of the sheet is
\begin{equation}\label{eq:H}
H_\pm(\xi,v) =\frac{\pm \cot (\theta/2) \csc\gamma_{\pm}}{2r\left[\sin \gamma_\pm \mp v (1 \pm \gamma'_\pm)\right]},
\end{equation}
where  $(.)'=d/d\xi(.)$. Then the energy of each surface $E_b = B \int_0^{2 \pi} \int^{v^{max}_\pm}_0 H_\pm^2 dv d\xi$, where $B$ is the bending stiffness of the material of the sheet. Carrying out the  integral along the generators, $v$, explicitly leads to the following scaled bending energy for the two surfaces
\begin{eqnarray}\label{eq:bending2}
\frac{E_{b}}{B} &=& \frac{1}{8} \int_0^{2\pi}d\xi \sum_\pm  \frac{ \cot^2(\theta/2) \csc^2 \gamma_\pm}{1 \pm \gamma_\pm^{\prime}}\\
&&\times \ln \left[\frac{\sin\gamma_\pm}{\sin \gamma_\pm - v^{max}_\pm(\xi) \left(1 \pm \gamma_\pm^{\prime} \right) } \right].\nonumber
\end{eqnarray} 
We see that (\ref{eq:bending2}) is determined entirely in terms of the geometry of the crease. To model the fold itself, we use a phenomenological energy functional measuring the deviation of $\theta(\xi)$ from an equilibrium angle $\theta_0$, which we assume to be constant, so that the scaled crease energy
\begin{equation}\label{eq:phenom}
\frac{E_{c}}{B}=\frac{\sigma}{2} \int_0^{2 \pi} d\xi \left[\cos\left(\frac{\theta(\xi)}{2}\right)-\cos\left(\frac{\theta_0}{2}\right)\right]^2,
\end{equation}
where $\sigma=Kr/B$ is the ratio of the crease stiffness $K$ and the bending stiffness $B$. This energy reduces to a simple quadratic expression in the difference $\theta-\theta_0$ when $\theta \sim \theta_0$; although the precise form of this term does not affect our analytic results, it conforms to our numerical model \cite{stanfordcloth}.

The equilibrium shape of the curved crease results from  minimizing $E = E_{b}+E_{c}$  and is characterized by three parameters: the scaled natural width of the ribbon $\omega$, the natural dihedral angle between the two surfaces adjoining the crease $\theta_0$ and the dimensionless crease-surface energy scale $\sigma$, subject to appropriate boundary conditions. For example, an open circular crease has free ends and thus prefers to remain planar with $\tau=0$ since non-planarity  would increase both the curvature and torsion \cite{Note1}. A closed crease, however, is frustrated by geometry, forcing it to buckle, a fact that follows from the inequality $\kappa = 1/\sin(\theta/2) > 1$ when $\theta<\pi$ which requires $\int d\xi ~\kappa > 2 \pi$, and is incompatible with a planar crease with $\tau=0$ \cite{Fuchs and Tabachnikov99}. 

%%%%%%%%%%%%%%%%%%%%%%%%%
\begin{figure}[h]
\includegraphics[width=3in]{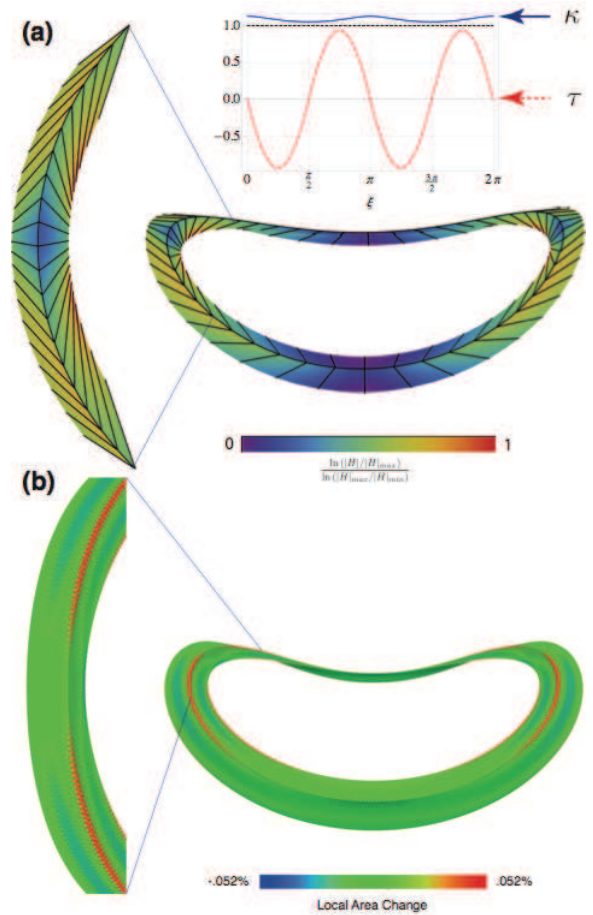}
\caption{\label{fig:fold9} (a) Perturbative fold of width $\omega=0.1$ and $\sigma=2/\sqrt{3}$ shaded by mean curvature. The generators are indicated by the lines on the surface. The inset shows the dimensionless torsion and curvature of the crease. (b) A simulated fold of width $\omega \approx .0994$ shaded by local area change relative to the flat state.
%(positive area change is red, negative area change is blue and no area change is green).
%The inset highlights the location of a unique area change pattern which emerges from the avoidance of singularities in the simulation. The other four regions of high area change correspond to twisted portions of the annulus with high fold torsion.
}
\end{figure}
%%%%%%%%%%%%%%%%%%%%%%%%%%%%%%

Though geometrical constraints induce buckling, the resulting fold shapes are determined by minimizing the total elastic energy consisting of contributions from the sheet (\ref{eq:bending2}) and the fold (\ref{eq:phenom}), expressed entirely in terms of the curvature and torsion of the crease \cite{CapovillaChryssomalakosGuven02,StarostinHeijden09}. For relatively narrow, but stiff, folds \textit{i.e.} $\omega \ll 1$ and $\sigma \gg 1$ that are weakly folded, i.e. so that the dihedral angle $\theta_0 \sim \pi$, and thence $\epsilon \equiv 1/\sin(\theta_0/2) - 1 \ll 1$.
%Expanding the scaled curvature and torsion of the fold to leading order in the parameters $1/\sigma, \epsilon, \omega$, we write $\kappa(\xi)=\kappa_0 + \delta\kappa(\xi)$ and $\tau(\xi) =\delta\tau(\xi)$ \cite{CapovillaChryssomalakosGuven02}, where $\kappa_0$ is the scaled equilibrium curvature of a planar circularly curved fold obtained by cutting the buckled fold. 
Then, we find that the total scaled energy $E=(E_b+E_c)/B$ simplifies to \cite{Note1}
\begin{equation}\label{eq:energyappr}
E \approx \int_0^{2 \pi} d\xi~\left\{ \frac{\sigma}{4 \epsilon} \left(\kappa- 1-\epsilon \right)^2+ \frac{\omega}{2} \tau^2\right\},
\end{equation}
in terms of the scaled curvature $\kappa$ and torsion $\tau$.
%\begin{equation}\label{eq:energyappr}
%\frac{E}{B} \approx \frac{1}{2}\int_0^{2 \pi} d\xi \left[\frac{\sigma}{\epsilon }  \delta \kappa^2+\frac{\omega  }{2\epsilon }\delta \kappa '^2+\omega  \delta \tau^2\right],
%\end{equation}
%where $4\sigma \approx - \omega \kappa_0^4/\left[1- \kappa_0/(1+\epsilon)/\sqrt{\kappa_0^2-1}\right]$.
We see that as $\sigma \rightarrow \infty$, the rescaled curvature $\kappa \rightarrow 1/\sin(\theta_0/2)=1+\epsilon$, the prescribed curvature. The minimal energy crease shape, therefore, minimizes $\tau^2$ subject to the constraints of fixed length and curvature. In this limit, the Euler-Lagrange equations become $[\tau''+ (1+\epsilon)^2 \tau]' \approx 0$ at constant curvature \cite{Note1}. If $\epsilon=0$, corresponding to a dihedral angle $\theta_0=\pi$  the solution to these equation is infinitely smooth. Otherwise, a solution of continuity class $C^4$ may be obtained to these equations with $\kappa=1+\epsilon$ and oscillating torsion,
\begin{equation}
\delta \tau = \Bigg\{
\begin{tabular}{cc}
$\tau_0 \left[ 1-\frac{\cos [(\xi-\pi/2)(1+\epsilon)]}{\cos[(\pi/2) (1+\epsilon)]} \right],$ & $0 \le \xi \le \pi$\\
$-\tau_0 \left[ 1-\frac{\cos[(\xi-3 \pi/2) (1+\epsilon)]}{\cos [(\pi/2) (1+\epsilon)]}\right],$ & $\pi \le \xi \le 2 \pi$.
\end{tabular}
\end{equation}
The absolute magnitude of the torsion $\tau_0$ is then determined by the condition that the curved fold has arc length $2 \pi r$, and consistent with the four-vertex theorem for closed convex space curves, there are four points with vanishing torsion \cite{Sedykh94}.

To go further, we can carry out an asymptotic analysis of the Euler-Lagrange equations obtained by minimizing $E=E_b+E_c$ must be performed by expanding the shape of the crease around a planar curve of constant curvature, $\kappa_0$. Following \cite{CapovillaChryssomalakosGuven02,StarostinHeijden09}, we write $\kappa = \kappa_0 + \delta \kappa$ and $\tau=\delta \tau$ and compute the Euler-Lagrange equations. To lowest order, we obtain an algebraic expression determining the ideal curvature of the crease, $\kappa_0$ for arbitrary $\sigma$, $\epsilon$ and $\omega$ \cite{Note1}, the curvature of an incomplete or severed planar annular fold with zero torsion. To next order, we find that both the curvature and torsion oscillate; a typical analytical solution is shown in Fig. \ref{fig:fold9}a, with the inset showing the oscillating torsion vanishing at the extrema of curvature (see Fig. \ref{fig:fold2}). Here the overall amplitude of $\tau$ is chosen to close the curve, with $\theta(0)$ and $\theta(\pi/2)$ parametrizing the solutions \cite{Note1}.

These qualitative features are also confirmed by direct numerical minimization of the energy of a triangular mesh model for the curved origami structure in which each edge is treated as a linear spring and adjacent triangles in each sheet have collinear normals and , with the scaled ratio of the bending stiffness to the stretching stiffness $B/Sl^2 \approx 10^{-3}$. However, adjacent triangles across the crease prefer a fixed, non-planar dihedral angle \cite{stanfordcloth}. The presence of a small but finite extensibility of this model implies that our simulations relax the isometry of the folding process and thus allow us to capture how extension and shear arise in wide folds  (Fig. \ref{fig:fold9}b). We find that the extensional and shear strains typically localize where the mean curvature becomes large, consistent with our isometric analytic theory (shown in Fig. \ref{fig:fold9}a).

%%%%%%%%%%%%%%%%%%%%%%%%
\begin{figure}[h]
\includegraphics[width=3.4in]{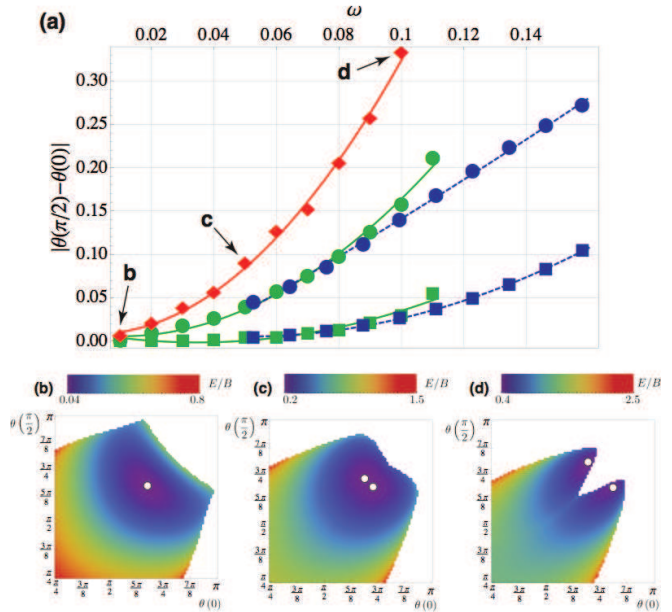}
\caption{\label{fig:fold6} (a) Angle differences $|\theta(\pi/2)-\theta(0)|$ as a function of $\omega$ with $\theta_0 = 2 \pi/3$. The red curve (diamonds) are computed from first-order perturbation theory with $\sigma=2/\sqrt{3}$ and $\theta_0=2 \pi/3$. The  numerical simulations (blue -- dashed lines) are shown with $\sigma=2 \sqrt{3}$ (circles) and $\sigma=160/\sqrt{3}$ (squares) and are compared with a nonperturbative variational ansatz (green -- solid line), $\kappa_{(2)}$ and $\tau_{(2)}$, described in the text with $\sigma=\sqrt{3}/40$ (circles) and $\sigma=\sqrt{3}/2$ (squares). Corresponding energy landscapes, as a function of $\theta(0)$ and $\theta(\pi/2)$ respectively, are shown for (b) $\omega=0.01$, (c) $0.05$, and (d) $0.1$, with energy minima drawn as white dots. }
\end{figure}
%%%%%%%%%%%%%%%%%%%%%%%%%%%%%%

Moving beyond the simple asymptotic theory for narrow folds, we consider the dependence of the solution on the scaled width by using the perturbative shapes as a variational \textit{ansatz} in the exact expression for the energy $E_b+E_c$. Since the shapes have a 4-fold symmetry, we expect to see a coincidence between $\theta(0)$ and $\theta(\pi/2)$. In Fig. \ref{fig:fold6}a, we plot $|\theta(0)-\theta(\pi/2)|$ for the minimal energy configuration as a function of the scaled width $\omega$. We see that when $\omega \lesssim 0.1$, annuli with large $\sigma$ have a nearly constant dihedral angle around the entire length of the fold, with $\theta(0) \approx \theta(\pi/2)$ for the narrowest fold widths. However, for small $\sigma$, the energy minimum generically has $\theta(0) \ne \theta(\pi/2)$; this discrepancy increases with  the scaled width $\omega$.  Plotting the corresponding energy landscape in Fig. \ref{fig:fold6}b-d for some representative values of $\omega$, we see that the energy contours develop forks because a range of $\theta(0)$ and $\theta(\pi/2)$ are forbidden by the geometric constraints that the generators of our two surfaces can intersect only outside the actual surface, else the bending energy diverges. 

To avoid the intersection of generators inside the outer surface requires
\begin{equation}\label{eq:bound1}
\gamma_+'  <  \frac{\sin\gamma_+}{v^{max}_+}-1\quad\mbox{and}\quad
\gamma_-' >  -\frac{\sin\gamma_-}{v^{max}_-}+1,
\end{equation}
which expression reduces to $\left|\tau'\right|<  \left(1-\omega\right)\cot \left(\theta/2\right)/\omega$, at points where $\tau=0$.
%\begin{equation}\label{eq:bound2}r^2 \left|\tau '(\xi^*)\right|<  \frac{1-\omega}{\omega}\cot \left(\frac{\theta(\xi^*)}{2}\right),\end{equation}
Similarly, to avoid the intersection of the generators on the inner surface inside the inner boundary requires the discriminant in equation (\ref{eq:vmax}) to be positive, implying a bound on the torsion,
\begin{equation}\label{eq:bound3}
\left|\tau+\frac{\theta'}{2}\right| <  \frac{1-\omega }{\sqrt{2\omega-\omega^2 }} \cot\left(\frac{\theta}{2}\right).
\end{equation} 
%If one thinks that the simplify version, constant curvature is more suitable for the paper, here it is:\begin{equation}r \left|\tau (\xi)\right| <  \frac{1-\omega }{\sqrt{2\omega-\omega^2}}\cot \left(\frac{\theta}{2}\right).\end{equation}
These geometrical bounds restrict the range of allowed torsion and thus the buckling of the crease. As a consequence, wide folds will become resistant to deformations as the sheet quickly reaches a regime where the generators start to nearly intersect in the neighborhood of $\xi=\pi/2$. In  Fig. \ref{fig:fold6}d, this is manifested by the presence of large forks carved out by the forbidden configurations. Since the energy minima occur close to the singularities, our perturbative expansion of the shape is approximate at best. However, even at intermediate widths, where the perturbative expansion should be at least qualitatively valid, the bifurcation of the minima show up in the shadows of the prominent forks observed in Fig. \ref{fig:fold6}d.
%In this way, the geometrical constraints shape the fundamental mechanical behavior of the folded annulus.

These calculations suggest a second improved \textit{ansatz}: $\kappa_{(2)} = \kappa_0 + \kappa_1 \cos(2 \xi)$ and $\tau_{(2)}=\tau_0 [\sin(2 \xi) +\eta \sin(4 \xi)]$, choosing $\tau_0$ to close the fold and $\eta$ to minimize the energy. When $\eta =0$ we now find very good agreement with the perturbative \textit{ansatz} previously considered. However, we find that $\eta \approx -0.45$ for large widths, which lowers the maximum of the torsion and better satisfies the singularity bounds in Eqs. (\ref{eq:bound1} - \ref{eq:bound3}). Using $\sigma$ as a fitting parameter, we see that $\theta(\pi/2)-\theta(0)$ agrees quite well with the numerical solutions for small $\omega$ and only diverges from numerical simulations for large widths, around $\omega\sim 0.08$ as shown in Fig. \ref{fig:fold6}a.

\begin{figure}[H]
\includegraphics[width=3.4in]{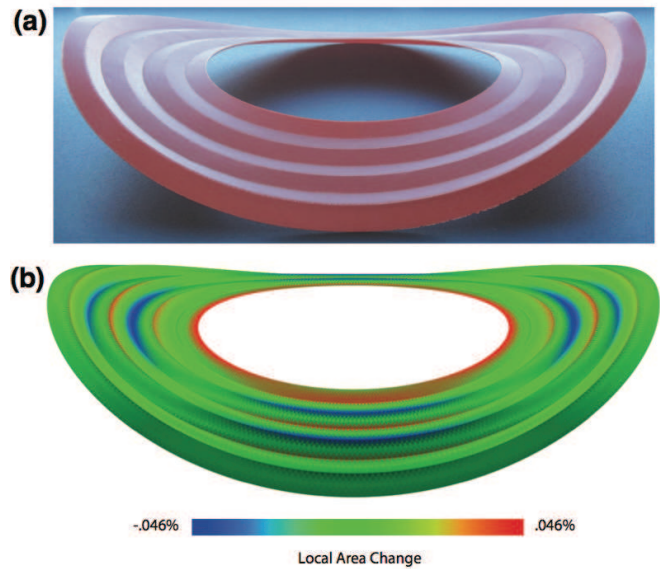}
\caption{\label{fig:fold8} (color online) (a) A plastic model of six circular folds generated by perforating the folds at equal intervals with a laser cutter. The ratio of outer to inner boundary is 2. (b) A simulation result with the same planar geometry as the plastic model in shaded by local area change. The multiple folds simulation has the same magnitude order of area change as that of the single fold simulation. }
\end{figure}

Finally, we consider structures built from multiple, concentric folds (Fig. \ref{fig:fold8}). Again, the large penalty for stretching leads to strong geometric constraints connecting the curvature and torsion of the crease to the dihedral angle of the fold given by equation (\ref{eq:constr1}), leading to self-similar creases and folds.  The generator on the outside of a crease must coincide with the generator on the inside of the next crease, fixing the angle $\gamma_+$ on the inside of the next crease, while the torsion determines the relationship between $\gamma_+$ and $\gamma_-$, so that the direction of the next set of generators emerges. This procedure follows from the first crease to the last crease until we reach a boundary or the generators cross. Our numerical simulations confirm this and further show that multiple creases have little stretching content and do buckle rigidly.

%%%%%%%%%%%%%%%%%%%%%%%%%%%%%%
%%%%%%%%%%%%%%%%%%%%%%%%%%%%%%

Our study of curved crease origami shows that a consequence of the fundamental frustration between folding along a curve and the avoidance of singularities and in-plane stretching imposes geometric constraints on the shape that are reflected in a bifurcation of the curvature of a closed crease of large width. Indeed, the coupling between shape and in-plane stretching endows these structures with a stiffness and response that is unusual, as we have demonstrated in the simplest of situations - a closed circular fold.  Moving forward, our approach may be generalized to more complex curves with variable dihedral angles in folded structures with curved creases and thus sets the stage for the analysis and design of these objects.

\begin{acknowledgements}
We thank Tom Hull and Pedro Reis for discussions and Badel Mbanga for helping with the photography. We acknowledge funding through NSF DMR 0846582, the NSF-supported MRSEC on Polymers at UMass (DMR-0820506) (MD, CS), the Wyss Institute for Bioinspired Engineering (LD, LM) and the MacArthur Foundation (LM).

\end{acknowledgements}

\bibliographystyle{apsrev}

\end{document}